\documentclass{emulateapj}
\usepackage{apjfonts}

\begin{document}

\title{The Extended Star Formation History of the Andromeda Spheroid
at 35 Kpc on the Minor Axis\altaffilmark{1}}

\author{
Thomas M. Brown\altaffilmark{2}, 
Rachael Beaton\altaffilmark{3}, 
Masashi Chiba\altaffilmark{4}, 
Henry C. Ferguson\altaffilmark{2},
Karoline M. Gilbert\altaffilmark{5}, 
Puragra Guhathakurta\altaffilmark{5}, 
Masanori Iye\altaffilmark{6}, 
Jasonjot S. Kalirai\altaffilmark{5,7},
Andreas Koch\altaffilmark{8},
Yutaka Komiyama\altaffilmark{6},
Steven R. Majewski\altaffilmark{3},
David B. Reitzel\altaffilmark{8}, 
Alvio Renzini\altaffilmark{9},
R. Michael Rich\altaffilmark{8}, 
Ed Smith\altaffilmark{2}, 
Allen V. Sweigart\altaffilmark{10},
Mikito Tanaka\altaffilmark{6}
}

\altaffiltext{1}{Based on observations made with the NASA/ESA {\it Hubble
Space Telescope}, obtained at STScI, and associated with proposal
10816.}

\altaffiltext{2}{Space Telescope Science Institute, 
Baltimore, MD 21218;  tbrown@stsci.edu, ferguson@stsci.edu, edsmith@stsci.edu} 

\altaffiltext{3}{Deptartment of Astronomy, University of Virginia, 
Charlottesville, VA 22904-4325; rlb9n@virginia.edu, srm4n@virginia.edu}

\altaffiltext{4}{Astronomical Institute, Tohoku University, Sendai 980-8578, 
Japan; chiba@astr.tohoku.ac.jp}

\altaffiltext{5}{University of California Observatories / Lick Observatory, 
University of California, Santa Cruz, CA 95064; 
kgilbert@ucolick.org, raja@ucolick.org, jkalirai@ucolick.org}

\altaffiltext{6}{National Astronomical Observatory, 
Tokyo 181-8588, Japan; m.iye@nao.ac.jp, komiyama@subaru.naoj.org,
mikito.tanaka@nao.ac.jp}

\altaffiltext{7}{Hubble Fellow}

\altaffiltext{8}{Department of Physics and Astronomy, 
University of California, Los Angeles, CA 90095;
akoch@astro.ucla.edu; reitzel@ucla.astro.edu, rmr@astro.ucla.edu}

\altaffiltext{9}{Osservatorio Astronomico, 
I-35122 Padova, Italy; alvio.renzini@oapd.inaf.it}

\altaffiltext{10}{NASA Goddard Space Flight Center, Greenbelt, MD
20771; allen.v.sweigart@nasa.gov}

\submitted{Accepted for publication in The Astrophysical Journal Letters}

\begin{abstract}

Using the {\it HST} ACS, we have obtained deep optical images reaching
well below the oldest main sequence turnoff in fields on the southeast
minor-axis of the Andromeda Galaxy, 35~kpc from the nucleus.  These
data probe the star formation history in the extended halo of
Andromeda -- that region beyond 30~kpc that appears both chemically
and morphologically distinct from the metal-rich, highly-disturbed
inner spheroid.  The present data, together with our previous data for
fields at 11 and 21~kpc, do not show a simple trend toward older ages
and lower metallicities, as one might expect for populations further
removed from the obvious disturbances of the inner spheroid.
Specifically, the mean ages and [Fe/H] values at 11 kpc, 21 kpc, and
35 kpc are 9.7~Gyr and $-0.65$, 11.0~Gyr and $-0.87$, and 10.5~Gyr and
$-0.98$, respectively.  In the best-fit model of the 35~kpc
population, one third of the stars are younger than 10~Gyr, while only
$\sim$10\% of the stars are truly ancient and metal-poor.  The
extended halo thus exhibits clear evidence of its hierarchical
assembly, and the contribution from any classical halo formed via
early monolithic collapse must be small.

\end{abstract}


\section{Introduction}

One of the primary quests of astronomy is measuring the
star formation history of giant galaxies.  The most direct tool for
such work is a color-magnitude diagram (CMD) reaching low-mass stars
below the main sequence turnoff.  The Advanced Camera for
Surveys (ACS) on the {\it Hubble Space Telescope (HST)} enabled the
application of this technique in the Andromeda Galaxy (M31),
the nearest giant galaxy to our own.  In a series of deep
surveys, we have been exploring the star formation histories in
M31's diverse structures (Brown et al.\ 2003, 2006, 2007).

The importance of M31 as a target for such studies is twofold.
First, M31 is the only giant spiral galaxy where we can resolve
the stellar main sequence with an external vantage point, thus
overcoming the large uncertainties due to reddening and distance that
can hamper studies of Milky Way field populations, most of which are
too distant for accurate parallaxes.  Second, recent
studies suggest that M31 is more representative of large
spiral galaxies than the Milky Way.  By comparing their locations in
the planes spanned by various parameters (rotational velocity,
absolute $K$ luminosity, disk angular momentum, [Fe/H] in the
outskirts, disk scale length, etc.), Hammer et al.\ (2007) argue that
M31 is more ``typical'' of large spirals than the Milky Way, which is
systematically offset by $\sim$1$\sigma$ from the dominant trends 
in these planes (see also
Flynn et al.\ 2006).  They attribute this distinction to an unusually
quiescent merger history in the Milky Way.  In contrast, 
the active merger history
in M31 is supported by star count maps (e.g.,
Ferguson et al.\ 2002; Ibata et al.\ 2007) showing a variety of
substructure, including a giant stellar stream (GSS) from a cannibalized
satellite (Ibata et al.\ 2001).

Going back to the work of Mould \& Kristian (1986), the M31
``halo'' was considered unusual because of its relatively high
metallicity ($\sim$10 times higher than that in the Milky Way halo).
Brown et al.\ (2003, 2006) have since demonstrated that these same
metal-rich populations were of intermediate age ($\sim$2--10~Gyr).
There is now strong evidence that the inner spheroid (within 15~kpc)
is polluted by the progenitor of the GSS; N-body
simulations (Fardal et al.\ 2007) and kinematical surveys (Gilbert et
al.\ 2007) demonstrate that much of the
substructure in the inner spheroid is due to debris dispersed from the
GSS's progenitor, while deep CMDs (Brown et
al.\ 2006) show strong similarities between the star formation
histories of the GSS and inner spheroid.

Recently, two groups independently discovered a vast extended halo in
M31 (Guhathakurta et al.\ 2005; Irwin et al.\ 2005), spanning
radii of 30--165~kpc and reaching lower metallicities similar to those
in the Milky Way halo (Kalirai et al.\ 2006; Koch et al.\ 2007).  
Our purpose here
is to measure the star formation history in this
extended halo and to compare it to that in the
interior fields.  We have used the surface brightness profile of
Guhathakurta et al.\ (2005) as a guide to this exploration (Figure 1).
Brown et al.\ (2003, 2006) explored a field at 11~kpc on the minor axis, a
metal-rich region of the spheroid that resembles a bulge, with
significant debris from the merger that produced the GSS.
Brown et al.\ (2007) investigated a field at 21~kpc, falling in the
transition zone between the inner and outer spheroid.  Here
we investigate two fields at 35~kpc, where the extended halo
begins to dominate.

\begin{figure}[t]
\epsscale{1.1}
\plotone{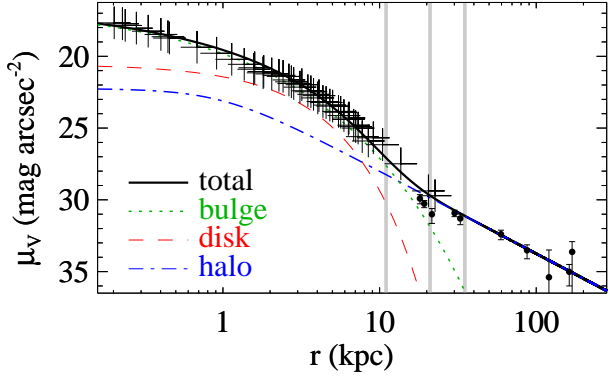}
\epsscale{1.1}
\caption{The minor-axis surface-brightness profile ({\it
    crosses}, Pritchet \& van den Bergh 1994; {\it error bars},
  Guhathakurta et al.\ 2005), along with one possible decomposition
  into disk, Sersic bulge, and $r^{-2.3}$ power-law halo (from
  Guhathakurta et al.\ 2005).  A clear break in the surface-brightness
  profile occurs at 20--30 kpc.  Our minor-axis fields ({\it grey
    shading}) sample the spheroid (i.e., bulge + halo) on either
  side of this transition region and within it.}
\label{proffig}
\end{figure}

\section{Observations and Data Reduction}

Using the ACS on the {\it HST}
we obtained deep optical images of two minor-axis
fields 2.6$^{\rm o}$ (35 kpc) from the M31 nucleus, at $\alpha_{2000}
= 00^h53^m28^s$, $\delta_{2000} = 39^{\rm
  o}49^{\prime}46^{\prime\prime}$ and $\alpha_{2000} = 00^h54^m08^s$,
$\delta_{2000} = 39^{\rm o}47^{\prime}26^{\prime\prime}$.  A third
field was also in the observing queue when the ACS Wide Field Camera
failed.  The use of three fields was intended to provide a sample in
the extended halo with at least as many stars as the single 21~kpc
field in the transition zone.  With only two fields, our 35~kpc
sample is smaller than desired, but still sufficient to characterize the
star formation history in the extended halo.

From Oct 2006 to Jan 2007, we obtained in each of our 35 kpc fields a
total of 8 hours of images in the F606W filter (broad $V$) and 13
hours in the F814W filter ($I$).  For one of
these fields, two visits in the F814W filter failed due to guide star
problems and were rescheduled with a 102$^{\rm o}$ change in orientation.
Every exposure was dithered to enable hot pixel removal,
optimal point spread function (PSF) sampling, smoothing of spatial
variation in detector response, and filling in the detector gap.  Our
reduction process is the same used by Brown et al.\ (2006) and only
briefly summarized here.  The images in a given field were registered,
rectified, rescaled to 0.03$^{\prime\prime}$ pixel$^{-1}$, and coadded
using the DRIZZLE package (Fruchter \& Hook 2002), with rejection of
cosmic rays and hot pixels.  PSF-fitting photometry, using the
DAOPHOT-II software of Stetson (1987), was corrected to agree with
aperture photometry of isolated stars, with the zeropoints calibrated
at the 1\% level.  The final catalog with the photometry of both
fields contains $\approx$8,500 stars (Figure 2).  Our photometry is in
the STMAG system: $m= -2.5 \times $~log$_{10} f_\lambda -21.1$.  For
those more familiar with the ABMAG system, ABMAG~=~STMAG~$-0.169$ for
$m_{F606W}$, and ABMAG~=~STMAG~$-0.840$ mag for $m_{F814W}$.  We
performed extensive artificial star tests to characterize the
photometric errors and completeness in the catalog.  To avoid
affecting the properties we were trying to measure, we added only 1000
artificial stars per pass, but by using thousands of passes, the tests
contain over 8 million artificial stars.

Spectroscopic surveys (Gilbert et al.\ 2007; Koch et al.\ 2007)
of red giant branch (RGB) stars in the M31 spheroid provide
kinematical context for the ACS images.  Figure 2 shows velocities
from regions that include and surround the ACS fields; for details on
targeting, data reduction, velocity fitting, and separation of
foreground dwarfs from M31 giants, see Gilbert et al.\ (2006) and
Guhathakurta et al.\ (2006).  Koch et al.\ (2007) obtain a similar
velocity distribution at 35~kpc.

The CMD of our 35~kpc field is shallower and far less crowded than the
CMDs of the inner spheroid, GSS, and outer disk obtained by
Brown et al.\ (2006).  Ideally, in our current program we would obtain
the same depth and star counts, but the scarcity of stars in these
fields forced us to investigate the trade-off between depth and star
counts that could be achieved in a program of reasonable size.
Through simulations, we found that we could determine the predominant
star formation history even with one-tenth the number of stars and 0.2
mag less depth.  The penalty is some loss of sensitivity to minority
population components (e.g., few percent ``bursts''), as demonstrated by
Brown et al. (2007).

The images at 21~kpc and 35~kpc suffer from charge transfer
inefficiency (CTI) in the ACS CCD due to radiation damage, but the
11~kpc images do not.  The difference in CTI is due to the
4-year baseline between the 11~kpc field and the later fields, and also
the higher stellar density in the 11~kpc field.  As in Brown
et al.\ (2007), we apply a CTI correction to our 35~kpc photometry.  The
correction is not applied to the artificial star tests, because
artificial stars are not clocked across the detector.  The
correction makes the faint stars brighter, but it has a small
effect on the star formation history fits (\S3); without the correction,
the 35~kpc population would appear to be 130 Myr older and 
0.11 dex more metal-poor.

\section{Analysis}

In Figure 2 we compare the spheroid populations at 11~kpc, 21~kpc, and
35~kpc, focusing on kinematics ({\it panels a--c}), CMDs ({\it panels
  d--f}), and star formation history fits ({\it panels g--i}).  Each
field hosts stars with a broad velocity distribution, centered on the
M31 systematic velocity.  None of the fields shows clear evidence of a
dominant, kinematically-cold component, as might be expected if there
were a majority contribution from a single stream or from the disk.

From CMD fitting, Brown et al.\ (2007) found the
population at 21~kpc to have a mean metallicity that is 0.2 dex lower and
a mean age that is 1.3~Gyr older than the population at 11~kpc.
Those findings were confirmed by visual inspection of the CMDs.
Compared to the 11~kpc population, the RGB ridge line of the 21~kpc
population is 0.016~mag bluer, the fraction of horizontal branch (HB)
stars blueward of the RR Lyrae gap is nearly twice as large, and the
RGB bump is 0.2~mag brighter (all implying lower metallicities), while
the luminosity difference between the HB and subgiant branch (SGB) is
0.12~mag larger (implying older ages).

If we perform the same fits and comparisons with the 35~kpc
population, we find that the populations at 21 and 35~kpc are distinct,
but the changes with increasing radius are not monotonic.
At $m_{F814W} = 27\pm 0.5$~mag, the median RGB color of
the 35~kpc population is only 0.004~mag bluer than that at 21~kpc
(identical within the uncertainties).  The RGB bump is not obvious
in the 35~kpc fields; a slight overdensity of stars immediately below
the red edge of the red clump may be due to the bump and clump
merging, which would be consistent with a lower [Fe/H] than that in
the 21~kpc field (where the RGB bump is bright but still distinct from
the HB).  All but 4 of the HB stars in the 35~kpc population fall in
the red clump, whereas one would expect twice as many blue HB stars 
if the fraction of such stars exceeded that in the 21~kpc
population.  Although HB morphology tends to become bluer at
decreasing [Fe/H], several other parameters
can affect the HB morphology; e.g., a younger age could compensate for
a lower [Fe/H] to put more stars in the red clump.  Comparing
the luminosity functions of each field, the luminosity difference
between the HB and SGB is $\sim$0.06~mag smaller in the 35~kpc fields
than in the 21~kpc field, implying a somewhat younger age at 35~kpc.
However, all of these visual inspections are hampered by the small
number statistics in the 35~kpc fields.

Next we turn to quantitative fitting of the 35 kpc CMD, using the
Starfish code of Harris \& Zaritsky (2001), following the methodology
of Brown et al.\ (2006).  As done previously, we fit the lower RGB,
SGB, and upper main sequence: $26.5 \leq m_{F814W} \leq 30$~mag and
$-0.9 \leq m_{F606W}-m_{F814W} \leq -0.1$~mag (but excluding a $0.1
\times 0.2$~mag region at the blue HB); the fit thus avoids parts of
the CMD with poor statistics, high foreground contamination, and
poorly-understood dependence on age and [Fe/H].  The isochrone library
comes from VandenBerg et al.\ (2006), 
transformed into the ACS bandpasses using the calibration of Brown et
al.\ (2005).  The age and metallicity
distributions in these fits are shown in the bottom panels of
Figure~2.  From the weights in Figures 2h and 2i, we find that the
$<$[Fe/H]$>$ is 0.1 dex lower and the $<$age$>$ is 0.5 Gyr younger in the
35 kpc field, compared to the 21~kpc field.

The 35~kpc population is inconsistent with a population that is
ancient and coeval; a third of the stars in the best-fit model are
younger than 10~Gyr.  If we exclude such stars from the fits, the
result is ruled out at 8$\sigma$ (where $\sigma$ is evaluated from repeated
draws on the best-fit model), but the exclusion of stars younger
than 6~Gyr is only ruled out at 2$\sigma$.  A fit excluding super-solar 
metallicities is within 1$\sigma$ of the best fit.
Attempts to fit the 21~kpc
and 35~kpc populations with the same star formation history imply they
are distinct at more than 3$\sigma$.  This is
not due to small-number statistics in the 35~kpc field. If we fit
100 random draws of 8500 stars from the 21~kpc CMD, in 95\% of the
Monte Carlo runs the $<$age$>$ is within 0.27~Gyr and the $<$[Fe/H]$>$
is within 0.07~dex of the values in the best fit to the full 21~kpc CMD.
If the best-fit models at 21 and 35~kpc
are normalized to the same stellar mass, only 44\%
of that mass occupies the same age and [Fe/H] space.

\section{Summary and Discussion}

We have measured the star formation history in three regions of the
M31 spheroid, spanning a large range in distance on the minor axis
(11~kpc, 21~kpc, and 35~kpc from the nucleus).  The [Fe/H]
decreases monotonically with increasing radius, with mean values of
$-0.65$, $-0.87$, and $-0.98$,
respectively.  However, the age does not monotonically increase
with increasing radius, having mean values of 9.68, 10.98, and 10.45~Gyr,
respectively.  

In their investigations of a metallicity gradient in the spheroid,
Koch et al.\ (2007) find that $<$[Fe/H]$>$ falls more rapidly with
radius than Kalirai et al.\ (2006).  Although our $<$[Fe/H]$>$ at
35~kpc is in good agreement with the trend of Kalirai et al.\ (2006)
and nearly 0.5 dex higher than that of Koch et al.\ (2007), the
latter study includes 15 stars in the vicinity of our ACS fields, and
for these, the $<$[Fe/H]$>$ agrees well with our value.  If the Koch
et al.\ (2007) trend is representative of the general gradient in the
spheroid, it would imply the ACS fields are in a local region of
systematically higher [Fe/H].  Note that our fit uses the lower RGB,
which is less sensitive to [Fe/H] than the upper RGB (used by Kalirai
et al.\ and Koch et al.), but it provides a larger RGB sample
and avoids the need to screen foreground dwarf contamination.

Our fits to the populations in the inner spheroid, the transition
zone, and the outer spheroid all require the presence of
intermediate-age stars (age $<$~10~Gyr).  Such populations are
expected if the spheroid formed via hierarchical
assembly.  This picture is also supported by recent star-count maps
(Ibata et al.\ 2007) that show several
stream-like structures crossing the minor axis beyond a distance of
40~kpc from the nucleus; relative to the bright ridge of stars in their 
Stream D (the nearest of these streams), our fields fall $\sim$30$^\prime$ 
(7 kpc) toward the interior, and may conceivably include 
debris from these streams.
Indeed, Bullock \& Johnston (2005) argue
that spheroids form inside-out, and that substructure should be
abundant in the outer regions of the spheroid.  A classical halo
that is ancient and metal poor, as might be expected from an early
monolithic collapse, does not comprise a large fraction of the
population at 35~kpc. If we arbitrarily define such an ancient
metal-poor population to be at [Fe/H]$<-1.5$ and age$\ge 12$~Gyr
(considered representative of the outer regions of the Milky Way
halo), the fraction of such stars in our best-fit model is only
$\sim$10\%.  Recent surveys (e.g., Ibata et al.\ 2007) imply it will
be difficult to find a truly ``clean''
region of the spheroid that is not criss-crossed by the
debris of mergers.

\acknowledgements

Support for GO-10816 is provided by NASA through a grant from
STScI, which is operated by AURA, Inc., under contract NAS
5-26555.  We acknowledge support from NSF grants
AST-0307966/AST-0507483 (PG), AST-0307931 (RMR),
AST-0307842/AST-0607726 (SRM, RLB), NASA/STScI grants
GO-10265/GO-10816 (PG, RMR), and NASA Hubble Fellowship grant
HF-01185.01-A (JSK).  We are grateful to P.\ Stetson for his DAOPHOT
code and to J.\ Harris for his Starfish code.

\clearpage

\begin{figure}[ht]
\epsscale{1.2}
%\epsscale{0.9}
\plotone{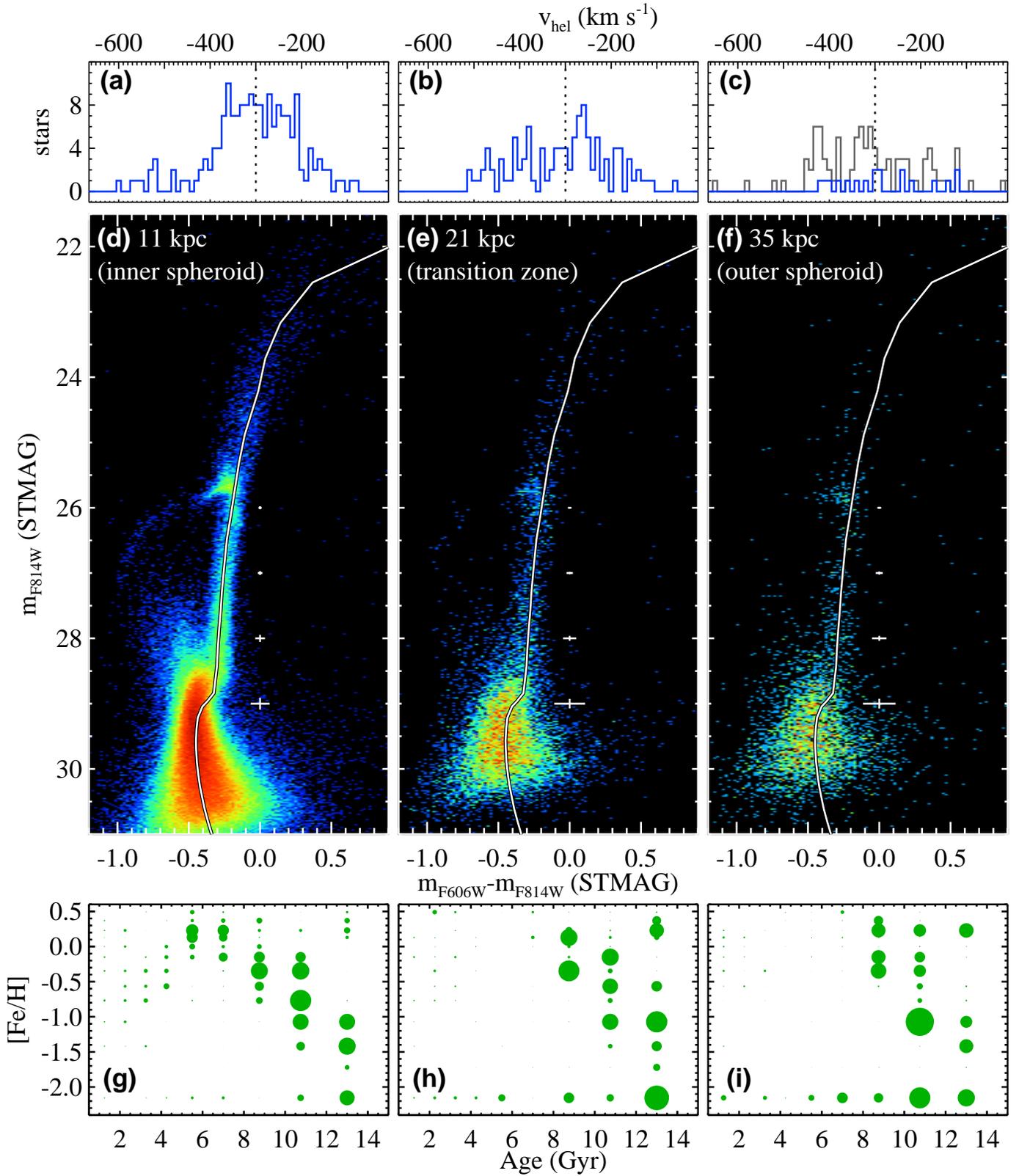} 
\epsscale{1.0}
\caption{{\it (a--c)} Velocities for RGB stars within 9$^\prime$ of
  our 11~kpc field, 14$^\prime$ of our 21~kpc field, and 17$^\prime$
  of our 35~kpc fields ({\it blue histograms}); these samples are
  taken from survey regions that include and surround our ACS fields.
  The velocities in each region exhibit a broad distribution near the
  M31 systemic velocity ({\it dashed line}).  To better demonstrate
  the broad (i.e., hot) distribution of velocities in the 35~kpc
  sample, we also show a dataset ({\it grey histogram}) that includes
  a region $\sim$30$^\prime$ to the West of our 35~kpc field; Gilbert
  et al.\ (2007) find no significant cold component in this region.
  {\it (d--f)} CMDs for the populations at 11~kpc, 21~kpc, and
  35~kpc. The ridge line of 47~Tuc ({\it white curve}; Brown et
  al.\ 2005), shifted to the M31 distance of 770~kpc (Freedman \&
  Madore 1991) and reddening of $E(B-V)=0.08$~mag (Schlegel et
  al.\ 1998) is shown for reference.  {\it (g--i)} The best-fit star
  formation histories for the populations at 11~kpc, 21~kpc, and
  35~kpc, from StarFish.  The area of each filled circle is
  proportional to the number of stars at that age and metallicity.
  The populations in all three fields exhibit broad ranges of
  metallicity and age, but the population at 11~kpc includes far more
  stars younger than 8~Gyr (due at least in part 
  to debris from the GSS).}
\end{figure}

\end{document}